\documentclass[11pt,a4paper]{article}
\usepackage{jheppub_kim}

\usepackage{pdflscape}
\usepackage{amsmath}
\usepackage{amssymb}
\usepackage{dcolumn}
\usepackage{bm}
\usepackage{color}
\usepackage{epsfig}
\usepackage{amsfonts}
\usepackage{graphicx}
\usepackage{subfigure}
\usepackage{dcolumn}

\newcommand{\be}{\begin{equation}}
\newcommand{\ee}{\end{equation}}
\newcommand{\bea}{\begin{eqnarray}}
\newcommand{\eea}{\end{eqnarray}}

\def\be{\begin{equation}}
\def\ee{\end{equation}}
\def\bea{\begin{eqnarray}}
\def\eea{\end{eqnarray}}

\begin{document}

\title{The Klein-Gordon Equation of a Rotating Charged Hairy Black Hole in (2+1) Dimensions}

\author{B. Pourhassan}

\affiliation{School of Physics, Damghan University, Damghan, Iran}

\emailAdd{b.pourhassan@du.ac.ir}

\abstract{In this paper, we consider the Klein-Gordon equation in a 3D charged rotating hairy black hole background to study behavior of a massive scalar field. In the general case we find periodic-like behavior for the scalar field which may be vanishes at the black hole horizon or far from the black hole horizon. For the special cases of non-rotating or near horizon approximation we find radial solution of Klein-Gordon equation in terms of hypergeometric and Kummer functions. Also for the case of uncharged black hole we find numerical solution of the Klein-Gordon equation as periodic function which may enhanced out of the black hole or vanish at horizon. We find allowed boundary conditions which yield to the identical bosons described by scalar field.}


\maketitle

\section{Introduction}
Exact solution of Einstein-Maxwell-scalar theory in (2+1) dimensions has been obtained which described a charged black hole with a scalar hair includes a negative cosmological constant which non-minimally coupled to gravity \cite{Xu-Zhao}, while exact solution of the Einstein-scalar-AdS theory with a non-minimal coupling yield to a rotating hairy black hole \cite{Zhao-Xu-Zhu}. It is found that the form of scalar potential strongly is depend on the interaction between gravity and the scalar field. Combination of both solutions represented by the Ref. \cite{Sadeghi-Pourhassan-Farahani} which includes a charged rotating hairy black hole in three dimensions. In the case of the charged hairy black hole, relation with a charged extremal
black hole as well as charged non-extremal black hole has been discussed.\\
Holographic description of such solutions has been considered to study Brownian motion \cite{Sadeghi-Pourhassan-Pourasadollah}, and found similar behavior with the asymptotic uncharged case in the low frequency limit. Thermodynamics and statistics of these black holes at special limits investigated by recent literatures \cite{Belhaj-Chabab-Moumni-Masmar-Sedra1,Xu-Zhao-Zou,Sadeghi-Farahani,Zou-Liu-Wang-Xu,Belhaj-Chabab-Moumni-Masmar-Sedra2}. For example Xu et al. \cite{Xu-Zhao-Zou} focused on some new 3D rotating black hole with minimal coupling to gravity and investigated Hawking-Page phase transition. Moreover, spectroscopy of the black hole has been shows that the circumference of the charged hairy black hole is discrete and depends on the black hole parameters \cite{Sadeghi-Farahani}.\\ Therefore, mentioned 3D hairy black hole \cite{sak1,sak2} was interesting subject of several issues due to relation with BTZ black hole and application in AdS/CFT correspondence.\\
Here we would like to solve Klein-Gordon equation in a charged rotating 3D black hole with a scalar charge. Solutions of the Klein-Gordon equation in several backgrounds has been obtained such as \cite{10,11}. For example in the Ref. \cite{12} the Klein-Gordon equation has been analyzed in the geometry of a rotating black hole and stability of a scalar field has been investigated. Also exact solutions of the Klein-Gordon equation in the Schwarzschild black holes has been studied \cite{13}.\\
The paper is organized as follows. In next section we give brief review of the 3D charged rotating hairy black hole. We discuss about thermodynamical properties of charged rotating solution with a scalar charge. Then, in section 3, we write the Klein-Gordon equation. Solutions of the Klein-Gordon equation discussed in section 4. Finally in section 7 we give conclusion and summarize our results.

\section{Black hole properties}
The 3D gravitational system with a non-minimally coupled scalar field and self coupling potential $V(\phi)$ described by the following action \cite{Zhao-Xu-Zhu},
\begin{equation}\label{B1}
S=\frac{1}{2}\int{d^{3}
x\sqrt{-g}[R-g^{\mu\nu}\nabla_{\mu}\phi\nabla_{\nu}\phi-\frac{R}{8}\phi^2-2V(\phi)-\frac{1}{4}F_{\mu\nu}F^{\mu\nu}]},
\end{equation}
which yield to the following line element \cite{Zhao-Xu-Zhu},
\begin{eqnarray}\label{B2}
ds^{2}=-f(r)dt^{2}+\frac{1}{f(r)}dr^{2}+r^{2}(d\psi+\omega(r)dt)^{2},
\end{eqnarray}
where \cite{Sadeghi-Pourhassan-Farahani},
\begin{equation}\label{B3}
f(r)\approx3\beta-\frac{Q^2}{4}+(2\beta-\frac{Q^2}{9})\frac{B}{r}-Q^2(\frac{1}{2}+\frac{B}{3r})\ln(r)+\frac{(3r+2B)^{2}a^{2}}{r^{4}}+\frac{r^2}{l^2},
\end{equation}
where $Q$ and $a$ denote the electric charge and rotation parameter respectively, also $l$ is related
to the cosmological constant, $\Lambda=-\frac{1}{l^{2}}$. $\beta$
is an integration constant which is depend on the black hole charge and mass via the following relation,
\begin{equation}\label{B4}
\beta=\frac{1}{3}(\frac{Q^2}{4}-M).
\end{equation}
Finally the scalar charge $B$ related to the scalar field as follow,
\begin{equation}\label{B5}
\phi(r)=\pm\sqrt{\frac{8B}{r+B}}.
\end{equation}
Also one can obtain,
\begin{equation}\label{B6}
\omega(r)=-\frac{(3r+2B)a}{r^{3}},
\end{equation}
and scalar potential given by,
\begin{equation}\label{B7}
V(\phi)\approx
\frac{2}{l^{2}}+\frac{1}{512}\left[\frac{1}{l^{2}}+\frac{\beta}{B^{2}}+\frac{Q^{2}}{9B^{2}}\left(1-\frac{3}{2}\ln(\frac{8B}{\phi^{2}})\right)\right]\phi^{6},
\end{equation}
in agreement with the \cite{Xu-Zhao}. It has been reported that the black hole horizon given by \cite{Sadeghi-Pourhassan-Farahani},
\begin{equation}\label{B8}
r_{h}^{2}=\frac{B}{3}\left(\frac{7}{6}-2\frac{M}{Q^{2}}\right)\left(-1+\sqrt{1+\frac{216a^{2}Q^{2}}{B(\frac{7}{6}Q^{2}-2M)^{2}}}\right).
\end{equation}
It may be useful to study thermodynamics properties of the black hole. A numerical study has been made by the Ref. \cite{Naji}. The entropy given by,
\begin{equation}\label{B9}
S=4\pi r_{h},
\end{equation}
and the black hole temperature obtained as,
\begin{eqnarray}\label{B10}
\pi T&=&\frac{r_{h}}{2l}-\frac{Q^{2}}{8}\frac{1}{r_{h}}+\frac{(144M-13Q^{2})B}{864r_{h}^{2}}-\frac{9a^{2}}{2r_{h}^{3}}\nonumber\\
&-&\frac{9Ba^{2}}{r_{h}^{4}}-\frac{4a^{2}B^{2}}{r_{h}^{5}}+\frac{BQ^{2}}{12}\frac{\ln{r_{h}}}{r_{h}^{2}}.
\end{eqnarray}
It is obvious that $a$ and $B$ increases value of the temperature while $Q$ decreases it. Also it has been argued that, for the appropriate value of the black hole parameters, there is stable black hole.

\section{Klein-Gordon equation}
In order to study behavior of a scalar field $\Psi$ in the gravitational field of a charged rotating hairy black hole we should solve the Klein-Gordon equation in a curved space-time, which is given by \cite{Bezerra et al},
\begin{equation}\label{KG1}
\left[\frac{1}{\sqrt{-g}}\partial_{\mu}(g^{\mu\nu}\sqrt{-g}\partial_{\nu})+\mathcal{M}^{2}\right]\Psi=0,
\end{equation}
where $g=-r^{4}\omega(r)^{2}$ is determinant of the metric given by the equation (\ref{B1}). Using non-zero component we have the following equation,
\begin{equation}\label{KG2}
\left[\partial_{t}g^{tt}\partial_{t}+\partial_{t}g^{t\phi}\partial_{\phi}+\frac{1}{\sqrt{-g}}\partial_{r}(g^{rr}\sqrt{-g}\partial_{r})
+\partial_{\phi}g^{\phi\phi}\partial_{\phi}+\partial_{\phi}g^{\phi t}\partial_{t}+\mathcal{M}^{2}\right]\Psi=0,
\end{equation}
where,
\begin{eqnarray}\label{KG3}
g^{tt}&=&\frac{1}{-f(r)+r^{2}\omega(r)^{2}},\nonumber\\
g^{rr}&=&-\frac{1}{-f(r)+r^{2}\omega(r)^{4}},\nonumber\\
g^{\phi\phi}&=&-\frac{f(r)}{(-f(r)+r^{2}\omega(r)^{4})r^{2}},\nonumber\\
g^{t\phi}&=&\frac{\omega(r)^{2}}{-f(r)+r^{2}\omega(r)^{4}}=g^{\phi t}.
\end{eqnarray}
In order to solve the equation (\ref{KG2}) we use separation method and assume \cite{Bezerra et al},
\begin{equation}\label{KG4}
\Psi=R(r)e^{im\phi}e^{-int},
\end{equation}
where $n$ denotes massive scalar field frequency of energy spectrum.
Substituting the equation (\ref{KG4}) in to the equation (\ref{KG2}) we can find,
\begin{eqnarray}\label{KG5}
\partial_{r}^{2}R(r)&+&\frac{f(r)-r^{2}\omega(r)^{4}}{\omega(r)r^{2}}\partial_{r}\left(\frac{\omega(r)r^{2}}{f(r)-r^{2}\omega(r)^{4}}\right)\partial_{r}R(r)\nonumber\\
&+&\left[(f(r)-r^{2}\omega(r)^{4})\mathcal{M}^{2}-2mn\omega(r)^{2}-\frac{f(r)}{r^{2}}m^{2}+\frac{f(r)-r^{2}\omega(r)^{4}}{f(r)-r^{2}\omega(r)^{2}}n^{2}\right]R(r)=0.\nonumber\\
\end{eqnarray}
In the next section we try to solve above equation to find $R(r)$, and therefore total wave function.
\section{Solutions}
In order to solve the complicated equation (\ref{KG5}) we assume special conditions to simplify equation. After all we try to solve general case of the equation numerically.
\subsection{Near horizon}
First of all, we consider massive scalar field near the horizon of a charged rotating hairy black hole ($f(r)\approx0$). There are two situations, first we assume $f^{\prime}(r)\approx0$, and rewrite the equation (\ref{KG5}) as follow,
\begin{equation}\label{S1}
\partial_{r}^{2}R(r)+\frac{18(r+B)}{r(3r+2B)}\partial_{r}R(r)
+\frac{(3r+2B)^{2}a^{2}}{r^{6}}\left[n^{2}-2mn-\frac{(3r+2B)^{2}a^{2}}{r^{4}}\mathcal{M}^{2}\right]R(r)=0.
\end{equation}
We will discuss about solutions in two separate ways.\\
If we switch off scalar charge ($B=0$), then solution of the equation (\ref{S1}) given by,
\begin{equation}\label{S2}
R(r\sim r_{h})=\frac{1}{r_{h}^{\frac{3}{2}}}\left[C_{1}WM(\frac{n^{2}-2mn}{4\mathcal{M}},\frac{5}{4},\frac{9\mathcal{M}a^{2}}{r_{h}^{2}})
+C_{2}WW(\frac{n^{2}-2mn}{4\mathcal{M}},\frac{5}{4},\frac{9\mathcal{M}a^{2}}{r_{h}^{2}})\right],
\end{equation}
where $WM$ and $WW$ are Whittaker$M(x,y,z)$ and Whittaker$W(x,y,z)$ functions which are written in terms of hypergeometric and Kummer functions as follows,
\begin{eqnarray}\label{S3}
WM(x,y,z)&=&e^{-\frac{z}{2}}z^{y+\frac{1}{2}}F([\frac{1}{2}+y-x], [1+2y], z),\nonumber\\
WW(x,y,z)&=&e^{-\frac{z}{2}}z^{y+\frac{1}{2}}KU([\frac{1}{2}+y-x], [1+2y], z),
\end{eqnarray}
where $F$ and $KU$ are hypergeometric and Kummer U functions respectively. However, the boundary condition $R(r\rightarrow\infty)$ required that $C_{2}\rightarrow0$ since $WM(r\rightarrow\infty)=0$, while $WW(r\rightarrow\infty)=\infty$. On the other hand $R(r\rightarrow0)$ tells that  $WM(r\rightarrow 0)=\infty$, so $R(r\rightarrow0)$ has infinite value as expected (scalar field fall in to the black hole), so value of $C_{1}$ is arbitrary. Therefore, the total solution may be written as,
\begin{equation}\label{S3-1}
\Psi=C_{1}e^{im\phi}e^{-int}\frac{WM(\frac{n^{2}-2mn}{4\mathcal{M}},\frac{5}{4},\frac{9\mathcal{M}a^{2}}{r_{h}^{2}})}{r_{h}^{\frac{3}{2}}}.
\end{equation}

\begin{figure}[h!]
 \begin{center}$
 \begin{array}{cccc}
\includegraphics[width=50 mm]{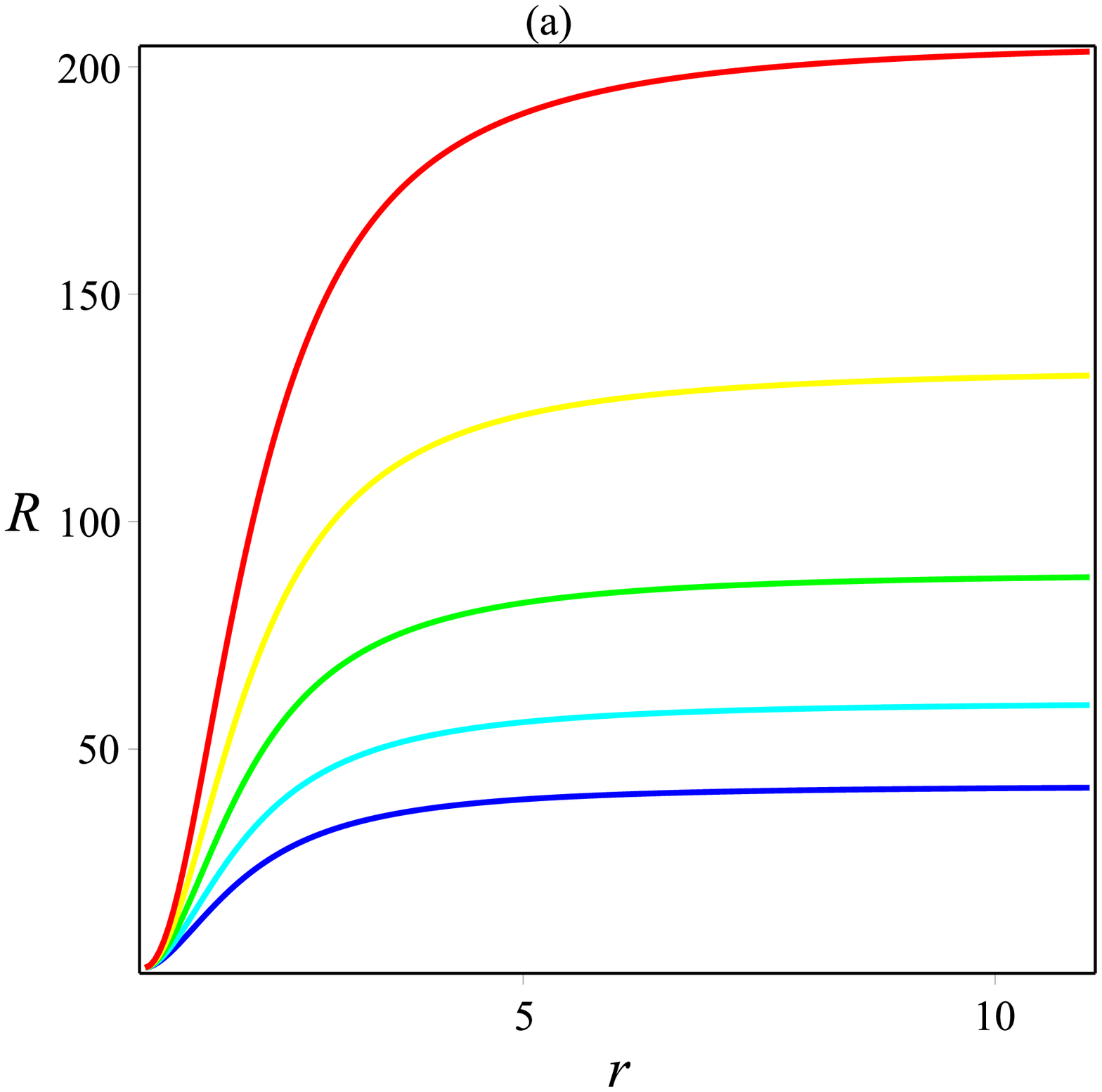}&\includegraphics[width=50 mm]{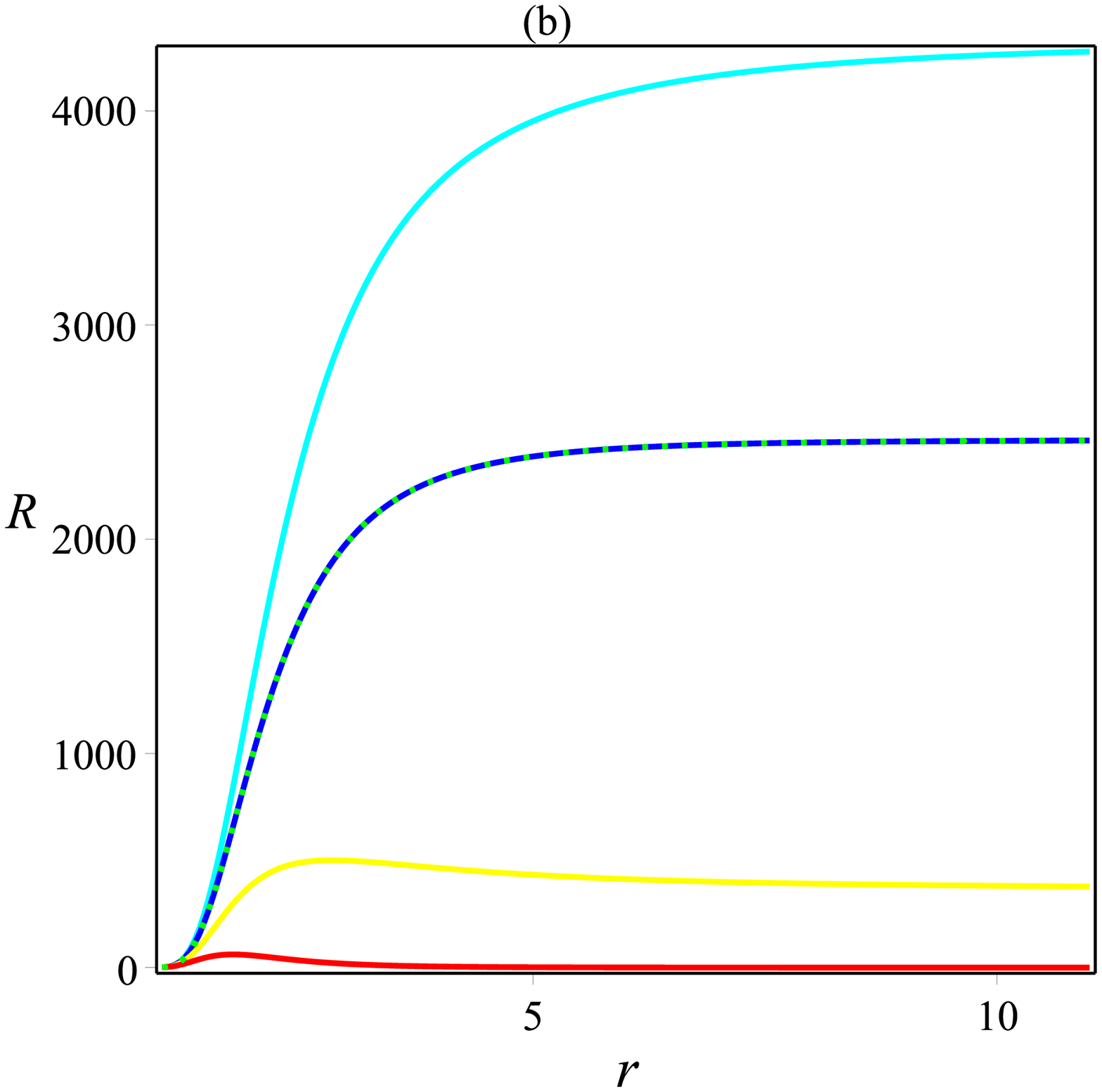}\\
\includegraphics[width=50 mm]{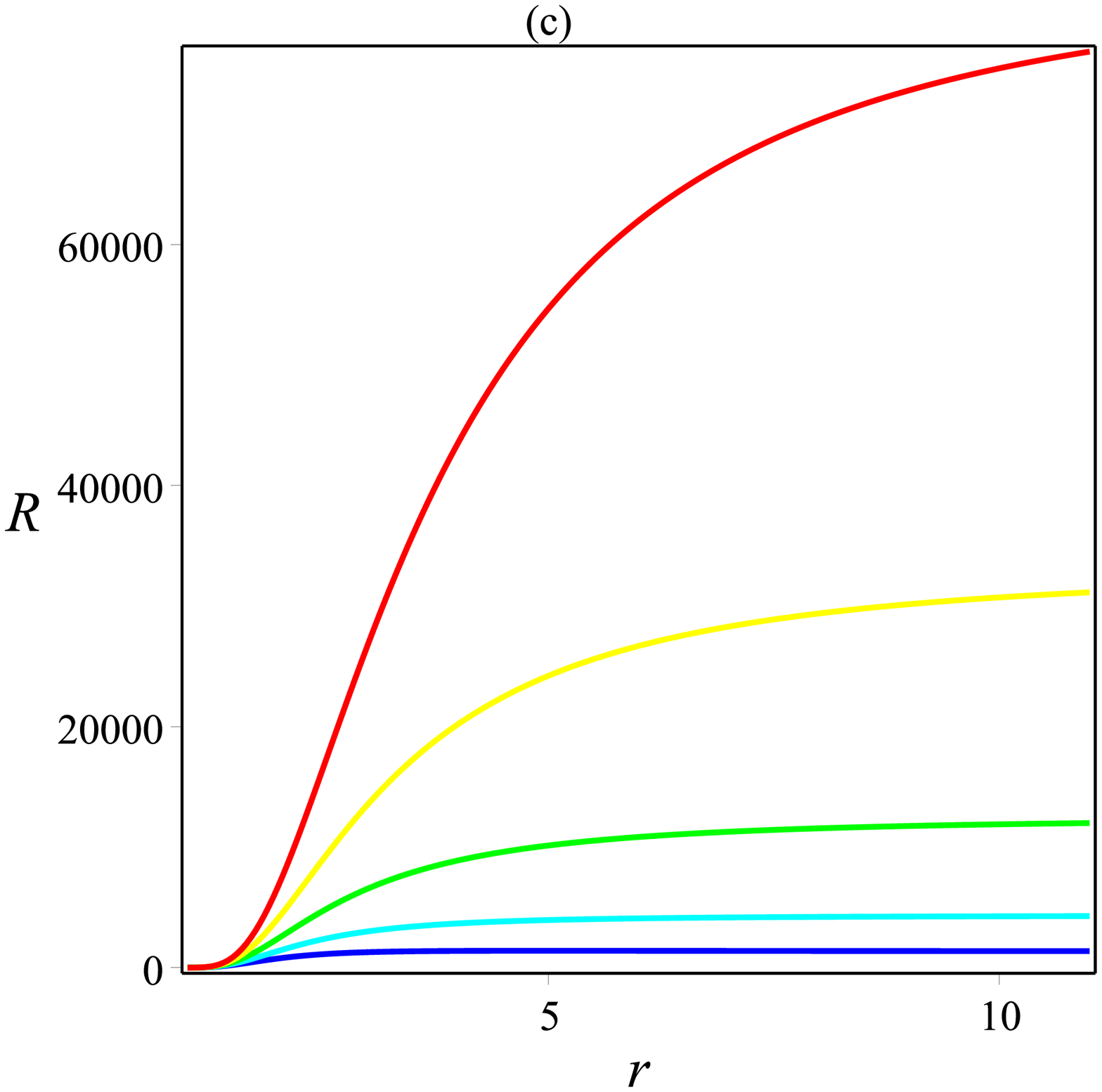}&\includegraphics[width=50 mm]{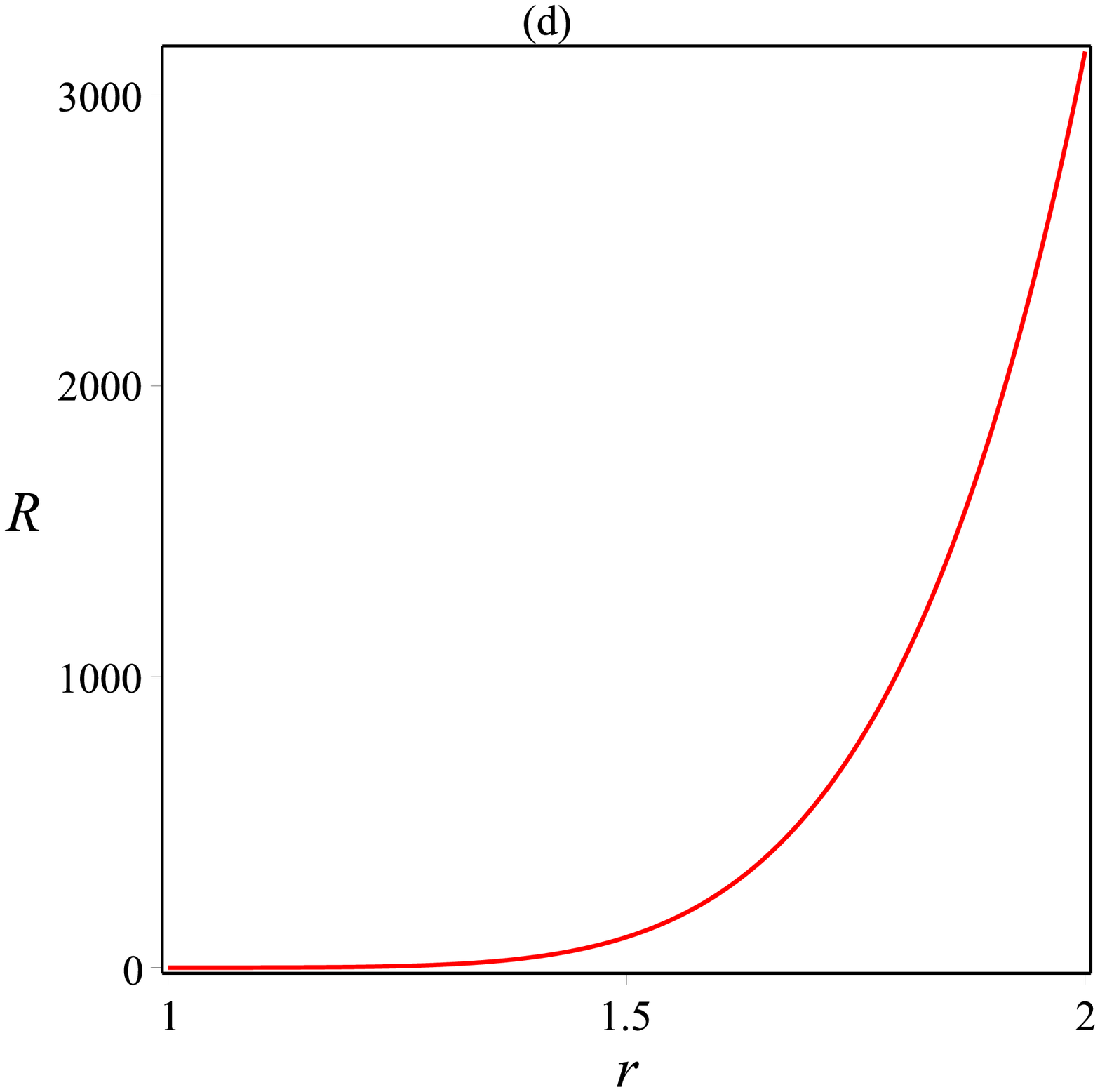}
 \end{array}$
 \end{center}
\caption{$R(r)$ in the case of near horizon for $\mathcal{M}=1$. (a) $a=m=n=1$, $B=0$ (blue), $B=0.1$ (cyan), $b=0.2$ (green), $B=0.3$ (yellow), $B=0.4$ (red). (b) $a=m=B=1$, $n=0$ (blue), $n=1$ (cyan), $n=2$ (green), $n=3$ (yellow), $n=4$ (red). (c) $a=B=n=1$, $m=0$ (blue), $m=1$ (cyan), $m=2$ (green), $m=3$ (yellow), $m=4$ (red). (d) the case of $f^{\prime}(r)\neq0$ with $\mathcal{M}=Q=B=m=n=M=l=1$.}
 \label{fig:1}
\end{figure}

Also, we can discuss numerically about the case of $B\neq0$. Our numerical results have shown by the Fig. \ref{fig:1}. We can see that increasing scalar charge as well as $m$ and $a$ increase value of $R(r)$. Situation is different for $n$. Increasing $n$ between $0\leq n\leq1$ increases value of $R$ while increasing $n$ for $n>1$ decreases value of $R$.\\
When we fixed all parameters, the value of $r_{h}$ depends only on $Q$, therefore we can say that the value of $R$ increases by the black hole charge to yield a finite value. We believed that results of this subsection will be useful to study particle acceleration \cite{Sadeghi-Pourhassan-Farahani,P1,P2} where near horizon solutions are important and may yield to infinite energy at the horizon of a rotating black hole.\\
The second case with $f(r)\approx0$ while $f^{\prime}(r)\neq0$ yields to the solution presented by the Fig. \ref{fig:1} (d). We can see completely opposite behavior, $R$ is increasing function of $r$ and may diverge exactly at the horizon.

\subsection{Non-rotating black hole}
If we assume $a=0$, then we recovered non-rotating black hole with the following Klein-Gordon equation,
\begin{equation}\label{S4}
\partial_{r}^{2}R(r)+\left[f(r)\mathcal{M}^{2}-\frac{f(r)}{r^{2}}m^{2}+n^{2}\right]R(r)=0.
\end{equation}
The simplest case of $B=Q=0$ gives again a solution in terms of Whittaker$M$ and Whittaker$W$ as follow,
\begin{eqnarray}\label{S5}
R(r)&=&\frac{C_{1}}{r^{\frac{1}{2}}}\left[WM(\frac{m^{2}-l^{2}n^{2}+M\mathcal{M}^{2}l^{2}}{4l\mathcal{M}},\frac{\sqrt{1-4\mathcal{M}m^{2}}}{4},\frac{\mathcal{M}r^{2}}{l})\right]\nonumber\\
&+&\frac{C_{2}}{r^{\frac{1}{2}}}\left[WW(\frac{m^{2}-l^{2}n^{2}+M\mathcal{M}^{2}l^{2}}{4l\mathcal{M}},\frac{\sqrt{1-4\mathcal{M}m^{2}}}{4},\frac{\mathcal{M}r^{2}}{l})\right].
\end{eqnarray}
Similar boundary condition as previous section suggests $C_{2}=0$ while $C_{1}$ is arbitrary.\\

\begin{figure}[h!]
 \begin{center}$
 \begin{array}{cccc}
\includegraphics[width=70 mm]{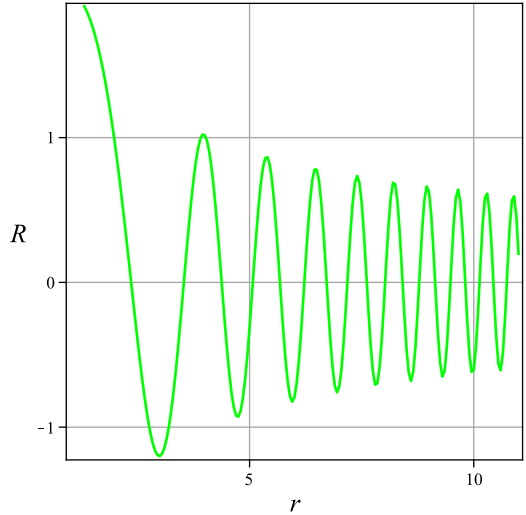}
 \end{array}$
 \end{center}
\caption{Typical behavior of $R(r)$ in the case of non-rotating for $\mathcal{M}=Q=B=m=n=M=l=1$.}
 \label{fig:2}
\end{figure}

For the general case of non-rotating solution we use numerical analysis which summarized in the Fig. \ref{fig:2}. We can see periodic like behavior which damped at large distance from black hole.\\
According to the selected values $B=Q=M=1$ and $a=0$ we can find that the scalar field wave function vanishes far from the black hole horizon. Fig. \ref{fig:2} suggests the following general function for the radial part,
\begin{equation}\label{fit1}
R(r)=c_{1}e^{-\alpha r}\sin(\frac{2\pi}{r_{0}}(r-r_{0})),
\end{equation}
where $r_{0}$, $c_{1}$ and $\alpha>0$ are arbitrary constants. Therefore, we can write total scalar field as,
\begin{equation}\label{psi1}
\Psi=c_{1}e^{-\alpha r}e^{im\phi}e^{-int}\sin(\frac{2\pi}{r_{0}}(r-r_{0})).
\end{equation}
It behaves as damped oscillator, and in fact it is bound state solution which satisfy the boundary condition being zero at infinity.

\subsection{Uncharged black hole}
\begin{figure}[h!]
 \begin{center}$
 \begin{array}{cccc}
\includegraphics[width=70 mm]{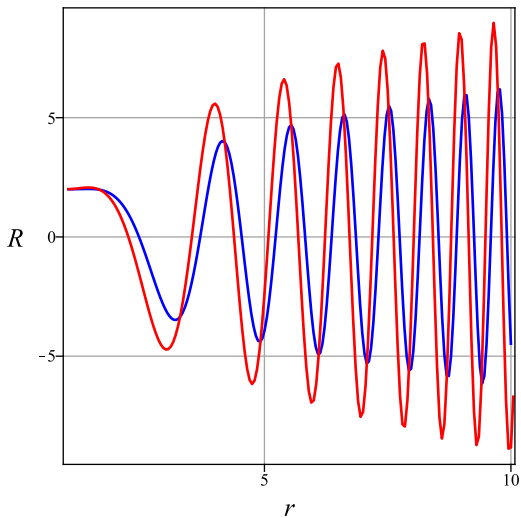}
 \end{array}$
 \end{center}
\caption{Typical behavior of $R(r)$ in the case of uncharged black hole for $\mathcal{M}=l=1$, $B=a=0.1$, $n=0$ and $M=2$. $m=1$ (blue),  $m=0$ (red).}
 \label{fig:3}
\end{figure}

If we assume $Q=0$ then the black hole horizon is root of the following equation,
\begin{equation}\label{S6}
\frac{r^{4}}{l^{2}}-Mr^{2}-\frac{2}{3}MBr+9a^{2}=0.
\end{equation}
Now we can obtain numerical solution of the Klein-Gordon equation (\ref{KG5}) as illustrated by the Fig. \ref{fig:3}. Unlike other cases we can see that periodic-like radial wave function enhanced out of horizon. With the selected value in Fig. \ref{fig:3}, outer horizon obtained as $r_{h}\approx1.5$, so we can see periodic behavior for scalar field outside of the black hole. We find solution outside of black hole for $n=0$ only, otherwise there is no solution outside and wave function vanishes at horizon.\\
Fig. \ref{fig:3} suggests the following general function for the radial part,
\begin{equation}\label{fit2}
R(r)=c_{2}e^{\gamma r}\sin(\frac{2\pi}{r_{0}}(r-r_{0})),
\end{equation}
where $r_{0}$, $c_{2}$ and $\gamma>0$ are arbitrary constants. Therefore, we can write total scalar field as,
\begin{equation}\label{psi2}
\Psi=c_{2}e^{\gamma r}e^{im\phi}e^{-int}\sin(\frac{2\pi}{r_{0}}(r-r_{0})).
\end{equation}
It seems it is unphysical solution; $R(r\rightarrow\infty)=\infty$, $R(r\rightarrow0)\simeq0$ which is opposite of the expected boundary condition.

\subsection{General case}
\begin{figure}[h!]
 \begin{center}$
 \begin{array}{cccc}
\includegraphics[width=70 mm]{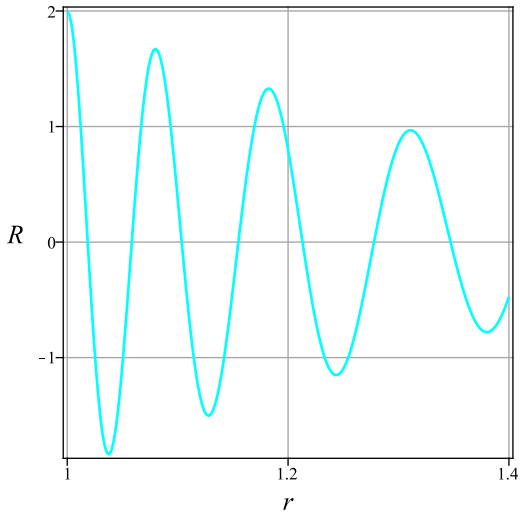}
 \end{array}$
 \end{center}
\caption{Typical behavior of $R(r)$ in the case of non-rotating for $\mathcal{M}=B=M=l=a=1$ and $Q=m=n=2$.}
 \label{fig:4}
\end{figure}

Numerically, we can solve general equation given by (\ref{KG5}) to obtain behavior of $R(r)$, hence scalar field $\Psi$. In the Fig. \ref{fig:4} we can see periodic-like behavior which vanished far from the black hole.\\
According to the relation (\ref{B8}) and selected values of the black hole parameters in the Fig. \ref{fig:4} we can find that the scalar field wave function vanishes at horizon. Fig. \ref{fig:4} suggests a general solution like (\ref{psi1}). Therefore, scalar field behaves as damped periodic oscillator vanished at horizon. It is indeed bound state which vanishes at infinity. Our solution describe identical bosons with the fitted function of the form (\ref{fit1}). It means that, effect of rotation on the solution of the Klein-Gordon equation is infinitesimal.

\section{Conclusion}
In this paper, we found analytical and numerical solutions of the radial part of the Klein-Gordon equation for a massive scalar field in the space-time of a rotating charged hairy 3D black hole. First of all we analyzed near horizon behavior and found that radial part of solutions are increasing function of $r_{h}$. Vanishing scalar charge or rotation parameter may give analytical solutions in terms of hypergeometric and Kummer functions. Our numerical analysis suggested periodic-like function for the radial part of the Klein-Gordon equation, which interpreted as identical bosons. We fitted functions corresponding to curves and obtained analytical solutions. In the general case it is damped far from black hole while in the case of uncharged black hole it is enhanced and grows with radius, which considered as un-physical solution. We found that the effect of rotation on the solution of the Klein-Gordon equation is infinitesimal.\\
For the future work, it is valuable to check the quantum singularities of rotating charged hairy 3D space-time \cite{unver}, also it is interesting to apply method of the Antoci and Liebscher \cite{AL} to the 3D rotating charged hairy black hole. Our results may be useful to study the black hole radiation or scattering process, also it is possible to use them to investigate stability of black holes.

\end{document}